\documentclass[prl,aps,superscriptaddress,twocolumn,amssymb,amsmath,longbibliography]{revtex4-2}
\usepackage{graphicx}
\usepackage{theorem}
\usepackage{dcolumn}
\usepackage{bm}
\usepackage{mathrsfs}
\usepackage{setspace}

\begin{document}
\title{Single-electron qubits based on quantum ring states on solid neon surface}

\author{Toshiaki Kanai}
\affiliation{National High Magnetic Field Laboratory, 1800 East Paul Dirac Drive, Tallahassee, Florida 32310, USA}
\affiliation{Department of Physics, Florida State University, Tallahassee, Florida 32306, USA}

\author{Dafei Jin}
\email[Email: ]{dfjin@nd.edu}
\affiliation{Department of Physics and Astronomy, University of Notre Dame, Notre Dame, Indiana 46556, USA}

\author{Wei Guo}
\email[Email: ]{wguo@eng.famu.fsu.edu}
\affiliation{National High Magnetic Field Laboratory, 1800 East Paul Dirac Drive, Tallahassee, Florida 32310, USA}
\affiliation{Mechanical Engineering Department, FAMU-FSU College of Engineering, Tallahassee, Florida 32310, USA}

\date{\today}

\begin{abstract}
Single electrons trapped on solid neon surfaces (eNe) have recently emerged as a promising platform for charge qubits. Experimental results have revealed their exceptionally long coherence times, yet the actual quantum states of these trapped electrons, presumably on imperfectly flat neon surfaces, remain elusive. In this paper, we examine the electron's interactions with neon surface topography, such as bumps and valleys. By evaluating the surface charges induced by the electron, we demonstrate its strong perpendicular binding to the neon surface. The Schr\"{o}dinger equation for the electron’s lateral motion on the curved 2D surface is then solved for extensive topographical variations. Our results reveal that surface bumps can naturally bind an electron, forming unique quantum ring states that align with experimental observations. We also show that the electron's excitation energy can be tuned using a modest magnetic field to facilitate qubit operation. This study offers a leap in our understanding of eNe qubit properties and provides strategic insights on minimizing charge noise and scaling the system to propel forward quantum computing architectures.

\end{abstract}
\maketitle

The success of quantum computing relies on qubits with long coherence times and swift operation~\cite{hanson_2007_RevModPhys, de_leon_2021_Science}. Among various types of qubits, charge qubits are noted for their fast operation speeds, resulting from strong coupling to electric fields. However, traditional semiconductor and superconducting charge qubits face challenges with charge noise, which limits their coherence times to about 1~$\mu$s~\cite{chatterjee_2021_NatRevPhys, heinrich_2021_NatNanotechnol, stano_2022_NatRevPhys}. On the other hand, charge qubits composed of electrons bound to the surfaces of ultra-clean quantum fluids and solids are predicted to exhibit prolonged coherence times~\cite{platzman_1999_Science, dykman_2000_FortschrPhys, lyon_2006_PhysRevA, Jin_2020_QST,Kawakami-2023-PRApp}. Over the past two decades, significant advancements have been made in comprehending these systems~\cite{kawakami_2019_PhysRevLett, zou_2022_NewJPhys, yunusova_2019_PhysRevLett, bradbury_2011_PhysRevLett, sabouret_2008_ApplPhysRev, byeon_2021_NatCommun, collin_2002_PhysRevLett, Rees-2011-PRL, schuster_2010_PhysRevLett, Yang-PRX-2016, koolstra_2019_Nature, dykman_2000_FortschrPhys, dykman_2003_PhysRevB, monarkha_2007_JLowTempPhys, kawakami_2021_PhysRevLett}. In particular, in a recent breakthrough, electron-on-solid-neon (eNe) qubits were shown to achieve coherence times on the order of 0.1~ms, positioning them at the forefront of this endeavor~\cite{zhou_2022_Nature, zhou_2023_NatPhys}.

\begin{figure}[t!]
\centering
\includegraphics[width=0.95\linewidth]{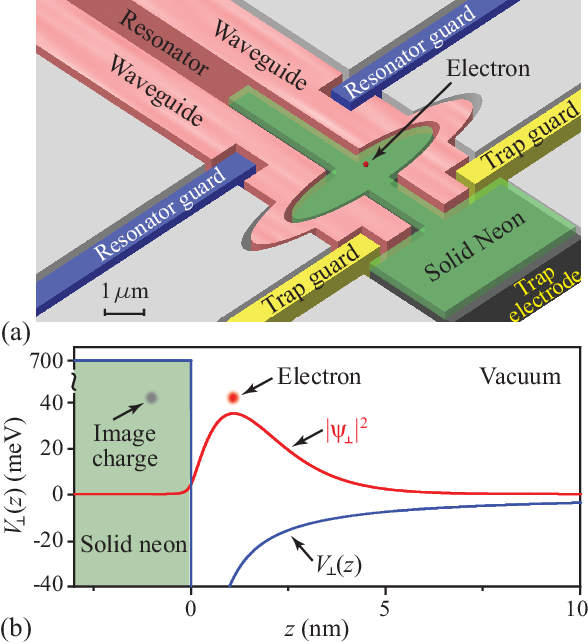}
\caption{(a) A schematic showing an electron trapped on the surface of solid neon in a superconducting microwave resonator. (b) Potential energy and the ground-state wavefunction of the electron near the flat solid-neon surface.}
\label{Fig1}
\end{figure}

The eNe qubit platform utilizes a hybrid quantum circuit structure consisting of a trap electrode placed inside a superconducting microwave resonator, accompanied by a number of guard electrodes, as shown schematically in Fig.~\ref{Fig1}(a)~\cite{zhou_2022_Nature, zhou_2023_NatPhys}. The trap electrode can be coated with a layer of solid neon. When an excess electron approaches the neon surface, the induced image charge in neon results in an attractive potential $V_{\bot}(z)=-[(\epsilon-\epsilon_0)/(\epsilon+\epsilon_0)]e^2/16\pi \epsilon_0 z$, where $\epsilon_0$ is the vacuum permittivity, $\epsilon=1.244\epsilon_0$ is the dielectric constant of solid neon, and $z$ is the vertical distance from the flat surface. On the other hand, due to Pauli exclusion between the excess electron and the atomic shell electrons, the solid neon appears as an energy barrier of about 0.7~eV to the electron \cite{cole_1969_PhysRevLett, cole_1970_PhysRevB, cole_1971_PhysRevB, jin_2020_QuantumSciTechnol}. These combined effects confine the electron in the $z$ direction with a ground-state wavefunction peaked at around 1~nm above the neon surface (see Fig.~\ref{Fig1}(b)). The ground-state energy is -15.8~meV, and the excitation energy to the first excited state in the $z$ direction is about 12.7~meV, equating to an activation temperature of 147~K. At typical experimental temperatures around 10~mK, the electron's motion perpendicular to the surface remains firmly in the ground state. When suitable voltages are applied to the guard electrodes, the electron's lateral motion can be confined within the elliptical trap region depicted in Fig.~\ref{Fig1}(a). By adjusting these voltages, one can align the transition from the electron's lateral ground state to its first excited state with the resonator's microwave photons for qubit operations \cite{zhou_2022_Nature}. The electron, hovering in a near-vacuum above the noble-element substrate, offers a qubit platform with minimal charge noise and hence exceptional coherence time. This qubit platform largely resolves the surface vibration and instability issues inherent in the electron-on-liquid-helium (eHe) qubit platform, which was proposed over two decades ago~\cite{platzman_1999_Science, dykman_2000_FortschrPhys} and has been extensively studied~\cite{yunusova_2019_PhysRevLett, bradbury_2011_PhysRevLett, sabouret_2008_ApplPhysRev, byeon_2021_NatCommun, collin_2002_PhysRevLett, Rees-2011-PRL, schuster_2010_PhysRevLett, Yang-PRX-2016, koolstra_2019_Nature, dykman_2000_FortschrPhys, dykman_2003_PhysRevB, monarkha_2007_JLowTempPhys, kawakami_2021_PhysRevLett}.

\begin{figure}[t!]
\centering
\includegraphics[width=1\linewidth]{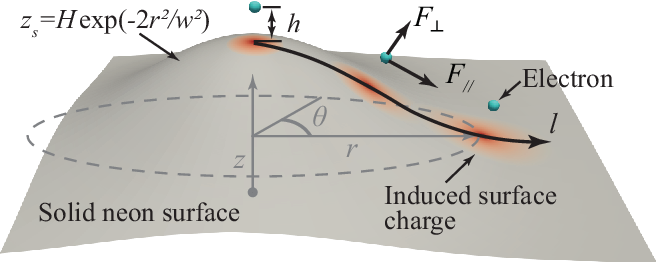}
\caption{A schematic showing the coordinate system adopted in our analysis of the electric forces acting on the electron near a curved solid neon surface.}
\label{Fig2}
\end{figure}

Nevertheless, recent experimental observations have revealed some intriguing behavior within the eNe qubit system. When the electric confining potential was reduced, the shift in the excitation spectrum associated with the electron's lateral motion was notably less than expected. Moreover, in some experimental runs, the electrons could remain anchored to the neon surface even after the confining potential was removed. These observations hint at the existence of alternative mechanisms affixing the electron on the neon surface. In this paper, we consider the interaction between the electron and deformations on the neon surface, such as bumps and valleys. These surface features may arise from the Stranski-Krastanov growth mode of neon at temperatures below its triple point~\cite{Esztermann-2002-PRL,Shchukin-2013-PRL}. Additionally, electrodes made of superconducting niobium deposited on silicon substrates inherently possess their own surface irregularities. Our analysis demonstrates that small surface bumps can capture the electron, resulting in unique quantum ring states that align with the experimental observations.

Without loss of generality, we consider a Gaussian bump on the neon surface as shown in Fig.~\ref{Fig2}, defined by the position vector $\vec{r}_s=\vec{r}_s(r_s,\theta_s,z_s)$, with $z_s=H\exp(-2r_s^2/w^2)$. Here, $H$ and $w$ denote the bump's height and width, respectively. To evaluate the interaction between an excess electron and this surface bump, the key is to determine the surface charge density $\sigma(\vec{r}_s)$ induced by the electron. To achieve this, we adopt an adaptive polar mesh around the electron's location to discretize the bump surface (see Supplemental Materials~\cite{Suppl}). For a surface element $\Delta S(\vec{r}_s)$, the continuity of the electric displacement across the surface requires \cite{jackson_1999_Book}: $\epsilon_0[\vec{E}_e+\Delta\vec{E}_s+\sum\limits_{\vec{r}_s'\neq\vec{r}_s}\Delta\vec{E}(\vec{r}_s')]\cdot \hat{n}_\bot=\epsilon[\vec{E}_e-\Delta\vec{E}_s+\sum\limits_{\vec{r}_s'\neq\vec{r}_s}\Delta\vec{E}(\vec{r}_s')]\cdot \hat{n}_\bot$
where $\vec{E}_e=\frac{-e}{4\pi\epsilon_0}\frac{\vec{r}_s-\vec{r}_e}{|\vec{r}_s-\vec{r}_e|^3}$ is the electric field produced at $\vec{r}_s$ due to the electron located at $\vec{r}_e$, $\Delta\vec{E}_s=(\sigma(\vec{r}_s)/2\epsilon_0)\hat{n}_\bot$ denotes the electric field generated by the induced charge at $\Delta S$ itself with $\hat{n}_\bot$ as the unit vector normal to $\Delta S$, and $\Delta\vec{E}(\vec{r}_s')=\frac{\Delta S(\vec{r}_s') \sigma(\vec{r}_s')}{4\pi\epsilon_0}\frac{\vec{r}_s-\vec{r'}_s}{|\vec{r}_s-\vec{r'}_s|^3}$ is the electric field produced at $\vec{r}_s$ by a surface element at $\vec{r'}_s$. This condition leads to the following expression for $\sigma_s(\vec{r}_s)$:
\begin{equation}
\begin{split}
\sigma(\vec{r}_s)=&\frac{1}{2\pi}\frac{\epsilon-\epsilon_0}{\epsilon+\epsilon_0}\bigg[-e\frac{(\vec{r}_s-\vec{r}_e)}{|\vec{r}_s-\vec{r}_e|^3}+\\
&\sum\limits_{\vec{r}_s'\neq\vec{r}_s} \Delta S(\vec{r'}_s)\sigma(\vec{r'}_s) \frac{(\vec{r}_s-\vec{r'}_s)}{|\vec{r}_s-\vec{r'}_s|^3} \bigg]\cdot \hat{n}_\bot
\end{split}
\label{Eq2}
\end{equation}
The above equation can be solved iteratively using an initial charge density $\sigma^{(0)}(\vec{r}_s)=\frac{-e}{2\pi}\frac{\epsilon-\epsilon_0}{\epsilon+\epsilon_0}\frac{(\vec{r}_s-\vec{r}_e)\cdot \hat{n}_\bot}{|\vec{r}_s-\vec{r}_e|^3}$, which is indeed the exact solution for a flat surface (see Supplemental Materials~\cite{Suppl}). In principle, for any given electron position $\vec{r}_e$, the associated surface charge density $\sigma(\vec{r}_s)$ should be determined. The electric potential energy of the electron can then be calculated as $V(\vec{r}_e)=\sum_{\vec{r}_s} \frac{1}{4\pi\epsilon_0}\frac{-e\Delta S(\vec{r}_s) \sigma(\vec{r}_s)}{|\vec{r}_s-\vec{r}_e|}$. This $V(\vec{r}_e)$, together with the 0.7-eV energy barrier from the curved neon surface, should be included in the Schr\"{o}dinger equation for the electron to solve for its eigen-wavefunctions in 3D space. However, this method is computationally intensive, limiting its feasibility for extensive investigations of diverse bump geometries.

\begin{figure*}[t]
\centering
\includegraphics[width=1\linewidth]{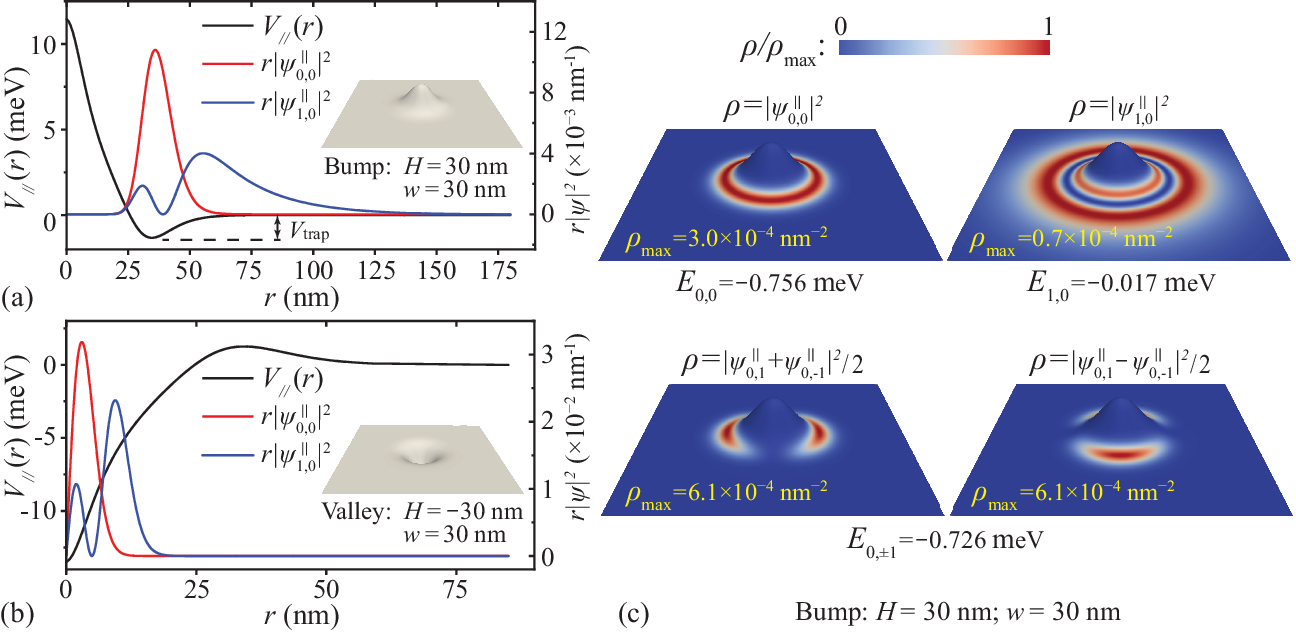}
\caption{(a) Potential energy profile associated with the electron's motion along a representative neon surface bump. The radial profiles of electron's wavefunction in its ground state and the first radially excited state are also shown. (b) Potential energy profile and the electron's radial wavefunction profiles along a representative neon surface valley. (c) 2D profiles of the ground state and excited states wavefunctions of the electron bound to a surface bump with $H=30$~nm and $w=30$~nm.
}
\label{Fig3}
\end{figure*}

Alternatively, we can calculate the electric force exerted on the electron due to the surface charge as $\vec{F}(\vec{r}_e)=\sum_{\vec{r}_s} \frac{-e\Delta S(\vec{r}_s) \sigma(\vec{r}_s)}{4\pi\epsilon_0}\frac{\vec{r}_s-\vec{r}_e}{|\vec{r}_s-\vec{r}_e|^3}$. From this, the forces parallel $\vec{F}_{\|}(\vec{r}_e)$ and perpendicular $\vec{F}_{\bot}(\vec{r}_e)$ to the local neon surface can be determined. For all the examined bump parameters $\{H,w\}$, we have found that $\vec{F}_{\bot}(\vec{r}_e)$ at any location above the curved neon surface only deviates minimally from the force on the electron from a flat surface (see Supplemental Materials~\cite{Suppl}). Therefore, the electron tends to bind at about 1~nm above the neon surface, exhibiting a distribution profile $|\psi_\bot(h)|^2$ perpendicular to the surface similar to that represented by the red curve in Fig.~\ref{Fig1}(b). This behavior is due to the bump surface's curvature radius being always substantially larger than 1~nm. Given this, we can focus on the electron's lateral motion along the curve neon surface, simplifying the 3D problem into a more manageable quasi-2D analysis.

Upon fixing the electron at a height $h$ above the neon surface, the potential energy associated with the electron's lateral motion can be calculated as $V_{\|}(l)=\int_l^\infty\vec{F}_{\|}(l')\cdot d\vec{l}'$, where $l$ is the coordinate mapped along the curved surface as shown in Fig.~\ref{Fig2}. Given the correlation between $l$ and $r$, we can represent $V_{\|}$ in terms of $r$. Fig.~\ref{Fig3}(a) shows the computed $V_{\|}(r)$ profile for the electron held at $h=1$~nm above a representative neon bump with $H=30$~nm and $w=30$~nm. Notably, $\vec{F}_{\|}$ changes sign from negative at large $r$ to positive at about $r=34$~nm, which results in a toroidal trapping potential encircling the bump with a potential depth $V_{trap}=-1.33$~meV. This phenomenon can be explained by considering how the distance between the electron and the surface varies with $l$, as elucidated in the Supplemental Materials~\cite{Suppl}. Considering the likehood $|\psi_\bot(h)|^2$ of the electron appearing at $h$, we have calculated $V_{\|}(r;h)$ at various $h$ and averaged the results as $\overline{V_{\|}}(r)=\int V_{\|}(r; h)|\psi_\bot(h)|^2dh$. It turns out that the averaged $\overline{V_{\|}}(r)$ profile shows little deviation from the curve displayed in Fig.~\ref{Fig3}(a), especially in the toroidal trap region (see Supplemental Materials~\cite{Suppl}). In subsequent analysis, we will fix the electron at $h=1$~nm for convenient exploration of various bump geometries.

To find the eigenstates associated with the electron's lateral motion, we solve the following Schr\"{o}dinger equation on the curved neon surface \cite{sakurai_2020_Book}:
\begin{equation}
\begin{split}
&E\psi_\|(r,\theta)=-\frac{\hbar^2}{2m_e}\nabla^2\psi_\|(r,\theta)+V_{\|}(r)\psi_\|(r,\theta)\\
&=-\frac{\hbar^2}{2m_er^2}\bigg[\frac{r}{h_r}\partial_r\left(\frac{r}{h_r}\partial_r\psi_\|\right)+\partial_\theta^2\psi_\|\bigg]+V_{\|}(r)\psi_\|,
\end{split}
\label{Eq3}
\end{equation}
where $m_e$ is the electron mass, $\hbar$ is the reduced Planck's constant, and $h_r=\sqrt{1+\frac{16Hr^2}{w^2}\exp(-\frac{4r^2}{w^2})}$ is the Lam\'e coefficient for the Gaussian surface \cite{thomas_1961_Book} (see Supplemental Materials for derivation~\cite{Suppl}). The wavefunction of an eigenstate can be expressed as $\psi^\|_{n_r,m_z}(r,\theta)=R_{n_r,m_z}(r)e^{im_z\theta}$, where $n_r$ and $m_z$ denote the radial and angular quantum numbers, respectively. Fig.~\ref{Fig3}(a) displays the radial profiles of both the ground state $\psi^\|_{0,0}$ and the first radially excited state $\psi^\|_{1,0}$ of the electron over a bump with $H=30$~nm and $w=30$~nm. The probability densities of the electron on the bump surface and the eigenenergies $E_{n_r,m_z}$ for these and two angularly excited states are shown in Fig.~\ref{Fig3}(c). Due to the geometry of the trapping potential, the eigenstates of the electron exhibit ring profiles around the bump.

We have also explored the interaction of the electron with valleys on the neon surface. Fig.~\ref{Fig3}(b) shows the potential energy profile $V_{\|}(r)$ for an electron bound on a Gaussian-shaped valley with $H=-30$~nm and $w=30$~nm, calculated using the same methodology applied for bumps. Given the inverse surface curvature as compared to the Gaussian bump shown in Fig.~\ref{Fig3}(a), the valley-associated potential is repulsive for $r\geq34$~nm. For $r<34$~nm, an axially symmetric potential well exists. The eigenstates of the electron confined in this potential well have been determined, with the radial profiles of the ground and the first radially excited states shown in Fig.~\ref{Fig3}(b). While such surface valleys can laterally confine the electron, they are not our primary focus. This is because when the electrons approach the solid neon, they adhere closely yet retain the ability to traverse the surface. From a distance, they are attracted to surface bumps but are repelled by valleys. Only upon exact positioning within the potential well region of the valleys, is there a possibility of them becoming confined. Therefore, our subsequent analysis will focus on surface bumps.

\begin{figure}[t]
\centering
\includegraphics[width=1\linewidth]{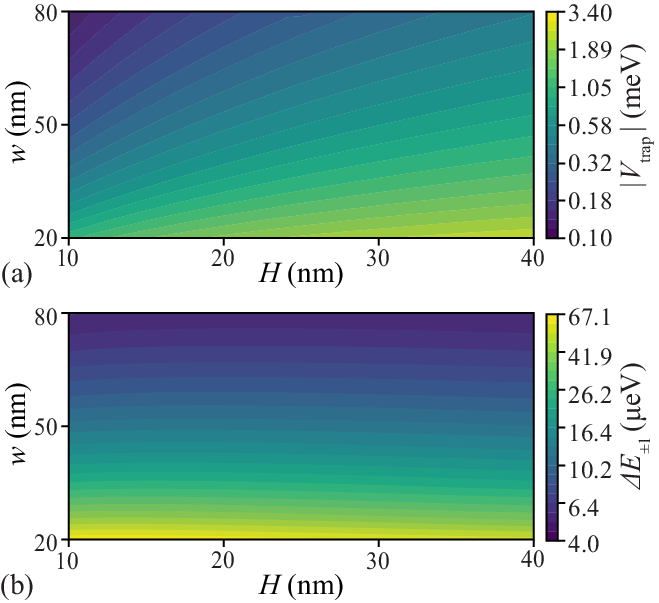}
\caption{(a) Trapping potential depth $|V_{trap}|$ and (b) excitation energy $\Delta E_{\pm1}=E_{0,\pm1}-E_{0,0}$ for the electron on neon surface bumps with various $H$ and $w$.}
\label{Fig4}
\end{figure}

\begin{figure}[t]
\centering
\includegraphics[width=1\linewidth]{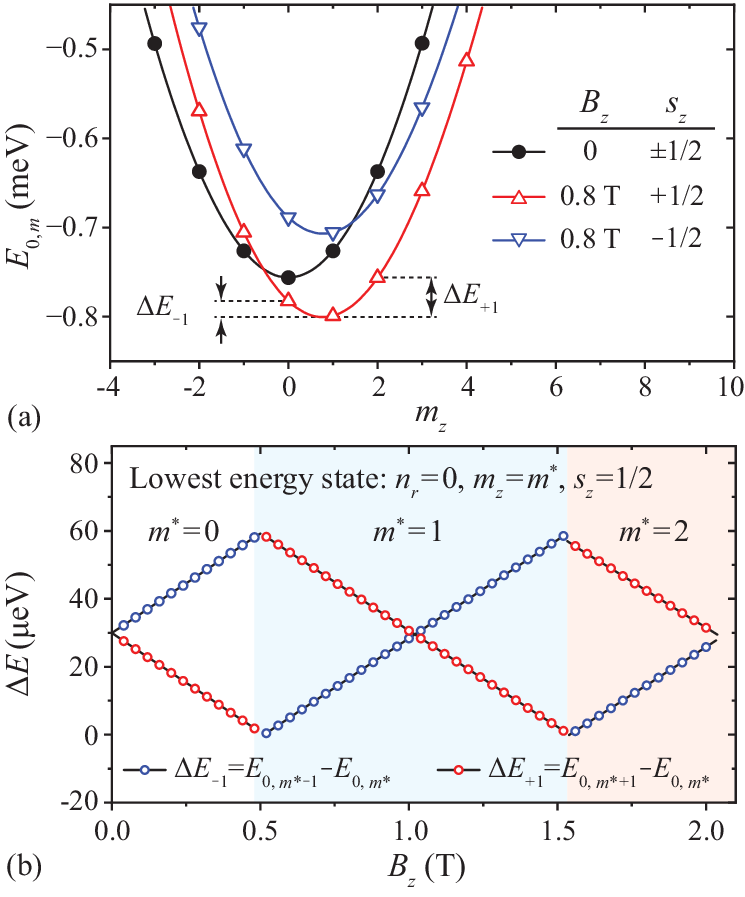}
\caption{(a) Eigenenergy versus angular quantum number $m_z$ for the electron on a bump with $H=30$~nm and $w=30$~nm. (b) Calculated excitation energy $\Delta E_{\pm1}$ from the lowest energy state at $m^*$ to the neighboring excited states at $m^*\pm1$ as a function of the applied magnetic field $B_z$.}
\label{Fig5}
\end{figure}

We have investigated a wide range of parameter combinations $\{H,w\}$ for neon surface bumps. In Fig.~\ref{Fig4}(a), the derived trapping potential depth $V_{trap}$ is presented as a function of $H$ and $w$. The magnitude $|V_{trap}|$ increases with increasing $H$ or decreasing $w$ and can reach a few meV, sufficient for electron confinement even in the absence of an applied electric trapping potential. Fig.~\ref{Fig4}(b) displays the calculated excitation energy $\Delta E_{\pm1}=E_{0,\pm1}-E_{0,0}$ for an optical transition from the ground state $\psi_{0,0}$ to the degenerate excited states $\psi_{0,\pm1}$. Since this transition occurs between the standing wave modes along the circumference of the toroidal trapping potential, $\Delta E_{\pm1}$ depends strongly on $w$ and is nearly independent of $H$. For bumps with $w$ of about 32~nm, $\Delta E_{\pm1}\simeq26$~$\mu$eV, matching well the energy of the microwave photons used in the experiments \cite{zhou_2022_Nature}.

The degeneracy of the states $\psi_{0,1}$ and $\psi_{0,-1}$, which is unfavorable for qubit operation, can naturally be lifted if the bump lacks perfect axial symmetry. Alternatively, one can apply a magnetic field to remove this degeneracy and to align the excitation energy precisely with the resonator's photons. For an applied uniform magnetic field $\vec{B}=-B_z\hat{z}$, the eigenenergy of the electron is given by $E^{(s_z)}_{n_r,m_z}=E_{r}+\frac{\hbar^2}{2m_e}\left\langle \frac{1}{r^2}\right\rangle m_z^2-\mu_BB_z(m_z+2s_z)$, where the radial energy contribution $E_r$ is mildly influenced by $m_z$ but strongly depends on $n_r$~\cite{sakurai_2020_Book} (see Supplemental Materials~\cite{Suppl}), $\left\langle \frac{1}{r^2}\right\rangle=\int2\pi rdr[R^2_{n_r,m_z}(r)/r^2]$, $\mu_B=\frac{e\hbar}{2m_e}$ is the Bohr magneton, and $s_z=\pm\frac{1}{2}$ is the spin quantum number. In the absence of the magnetic field ($B_z=0$), $E^{(s_z)}_{0,m_z}$ largely follows the trend $E^{(s_z)}_{0,m_z}-E^{(s_z)}_{0,0}\propto m_z^2$, with $\psi_{0,0}$ being the lowest energy state. Fig.~\ref{Fig5}(a) shows the computed $E^{(s_z)}_{0,m_z}$ versus $m_z$ for an electron confined on a bump with $H=30$~nm and $w=30$~nm. At finite $B_z$, the originally degenerate spin states now split. Furthermore, the linear term in $m_z$ in the eigenenergy expression can cause a shift of the $E^{(s_z)}_{0,m_z}$ curve as shown in Fig.~\ref{Fig5}(a). Given the considerable bump size, $\left\langle \frac{1}{r^2}\right\rangle$ is small and hence the shift caused by the linear term in $m_z$ can be significant, which may result in the lowest energy state with $m_z=m^*\neq0$. The excitation energies $\Delta E_{-1}=E_{0,m^*-1}-E_{0,m^*}$ and  $\Delta E_{+1}=E_{0,m^*+1}-E_{0,m^*}$ associated with the transitions from the new ground state at $m^*$ to the two neighboring excited states at $m^*\pm1$ now depend on $B_z$. In Fig.~\ref{Fig5}(b), we plot $\Delta E_{-1}$ and $\Delta E_{+1}$ as functions of $B_z$. The nearly linear dependence of $\Delta E_{\pm1}$ on $B_z$ suggests a robust capability of $B_z$ to precisely adjust the qubit's transition frequency.

Our theoretical findings may have profound implications for the design and optimization of eNe qubits. In previous experiments, the injected electrons could bind to naturally formed bumps of various sizes. Only when an electron bound to a bump of the correct size and located within the resonator cavity, did its states become manipulable by the cavity photons. Electrons bound to neon bumps of mismatched sizes would fail to resonate with the cavity photons. But these electrons could contribute to background charge noise, potentially leading to qubit decoherence, limiting the coherence times of eNe qubits to about 0.1~ms. Our research underscores the critical need for a systematic study of the neon growth process under various injection temperatures and pressures, aimed at optimizing the procedure for producing high-quality neon surfaces with minimal natural features. This effort is anticipated to significantly reduce the number of background electrons and hence enhance the coherence times of eNe qubits. On the other hand, one may intentionally fabricate bumps of the right size on the trap electrode to enhance the chances of trapping electrons for qubit operation. These bumps could be elongated in the direction of the cavity electric field to lift the degeneracy of the $\psi_{0,1}$ and $\psi_{0,-1}$ states and enhance the dipole coupling to the cavity electric field. When two or more such bumps are fabricated so that multiple electrons can be trapped simultaneously, these electrons' lateral states could be manipulated and entangled using the cavity photons for various multi-qubit gate applications.~\cite{Veldhorst-2015-Nature}.

It is also worth noting that the application of a non-uniform magnetic field parallel to the neon surface can lead to coupling between the electron's lateral motion and its spin degrees of freedom \cite{chen_2022_QuantumSciTechnol}. Such a concept has previously been proposed for the electron-on-helium system \cite{schuster_2010_PhysRevLett, zhang_2012_PhysRevB, dykman_2023_PhysRevB,Kawakami-2023-PRApp} and implemented in other systems like silicon qubits \cite{mi_2017_Science, mi_2018_Nature, samkharadze_2018_Science}. The resonator's photons can be employed to stimulate and control the electron's spin states. A recent estimation suggests that the spin coherence time of the eNe qubit could extend up to 81 s when using purified neon \cite{chen_2022_QuantumSciTechnol}. The potential of constructing a fault-tolerant quantum computer leveraging the spin states of eNe qubits presents a compelling avenue for further research and development.

\begin{acknowledgments}
T. K. and W. G. acknowledge the support by the National Science Foundation under Grant No. DMR-2100790 and the Gordon and Betty Moore Foundation through Grant GBMF11567. The work was conducted at the National High Magnetic Field Laboratory at Florida State University, which is supported by the National Science Foundation Cooperative Agreement No. DMR-2128556 and the state of Florida. D. J. acknowledges support from the Air Force Office of Scientific Research (AFOSR) under award No. FA9550-23-1-0636.
\end{acknowledgments}

\bibliography{Ref-neon}

\end{document}


\title{Supplemental Material for: Single-electron qubits based on ring-shaped surface states on solid neon}
\author{Toshiaki Kanai}
\affiliation{National High Magnetic Field Laboratory, 1800 East Paul Dirac Drive, Tallahassee, Florida 32310, USA}
\affiliation{Department of Physics, Florida State University, Tallahassee, Florida 32306, USA}

\author{Dafei Jin}
\affiliation{Department of Physics and Astronomy, University of Notre Dame, Notre Dame, Indiana 46556, USA}

\author{Wei Guo}
\email[Corresponding author: ]{wguo@magnet.fsu.edu}
\affiliation{National High Magnetic Field Laboratory, 1800 East Paul Dirac Drive, Tallahassee, Florida 32310, USA}
\affiliation{Mechanical Engineering Department, FAMU-FSU College of Engineering, Florida State University, Tallahassee, Florida 32310, USA}

\maketitle

\section{Surface charge density calculation}
\begin{figure}[b]
\includegraphics[width=1.0\linewidth]{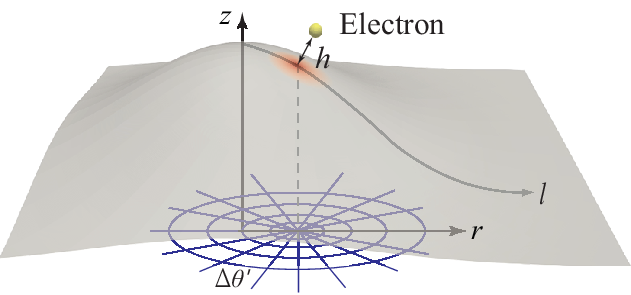}
\caption{Schematic illustration of the coordinate system and the adaptive mesh used for calculating the charge density on the neon surface.}
\label{Suppl-Fig1}
\end{figure}

As explained in the main paper, the surface charge density $\sigma(\vec{r}_s)$ at location $\vec{r}_s$ on the curved solid neon surface induced by an electron located at $\vec{r}_e$ can be expressed as:
\begin{equation}
\begin{split}
\sigma(\vec{r}_s)=&\frac{1}{2\pi}\frac{\epsilon-\epsilon_0}{\epsilon+\epsilon_0}\bigg[-e\frac{(\vec{r}_s-\vec{r}_e)}{|\vec{r}_s-\vec{r}_e|^3}+\\
&\sum\limits_{\vec{r}_s'\neq\vec{r}_s} \Delta S(\vec{r'}_s)\sigma(\vec{r'}_s) \frac{(\vec{r}_s-\vec{r'}_s)}{|\vec{r}_s-\vec{r'}_s|^3} \bigg]\cdot \hat{n}_\bot.
\end{split}
\label{SM-Eq1}
\end{equation}
In order to deduce the value of $\sigma(\vec{r}_s)$ from the above equation, the neon surface needs to be divided into discrete elements. Since $\sigma(\vec{r}_s)$ exhibits a high concentration around the electron's perpendicular projection on the curved surface, we first determine this projection point $\vec{r}_p=(r_p,0,h_p)$ for a given $\vec{r}_e$. Then, we employ an adaptive mesh that offers higher spatial resolution near the projection point. As shown in Fig.~\ref{Suppl-Fig1}, we create a polar mesh in the $z=0$ plane, with the origin set at $r=r_p$ and $\theta=0$. This mesh has an angular spacing of $\Delta \theta'=2\pi/N_{\theta'}$, where $N_{\theta'}$ is the number of angular grid points. The radial step length $\Delta r'$ is set to increase linearly with the distance $r'$ from the mesh origin as $\Delta r'=\Delta r_0+cr'$, where $\Delta r_0=10^{-3}$~nm and the coefficient $c=1.5\times10^{-3}$. Typically, we set the maximum range of this mesh to $r'=100$~nm with $N_r'$ grids in the radial direction. This adaptive mesh, once designed on the $z=0$ plane, is then mapped onto the curved neon surface so that a robust analysis of $\sigma(\vec{r}_s)$ can be made. The surface charge beyond the range covered by this mesh is negligible.

By inserting an initial charge density $\sigma^{(0)}(\vec{r}_s)=\frac{-e}{2\pi}\frac{\epsilon-\epsilon_0}{\epsilon+\epsilon_0}\frac{(\vec{r}_s-\vec{r}_e)\cdot \hat{n}_\bot}{|\vec{r}_s-\vec{r}_e|^3}$ into the right side of Eq.~\ref{SM-Eq1}, an updated charge density can be obtained. This new value can be re-inserted into the equation to further refine the charge density. The iteration is terminated once the relative change in the surface charge density between consecutive iterations, defined as $|\sigma^{(i+1)}(\vec{r}_s)-\sigma^{(i)}(\vec{r}_s)|/\sigma^{(i)}(\vec{r}_s)$, becomes less than $10^{-10}$ at all $\vec{r}_s$.

\begin{figure}[t]
\includegraphics[width=0.95\linewidth]{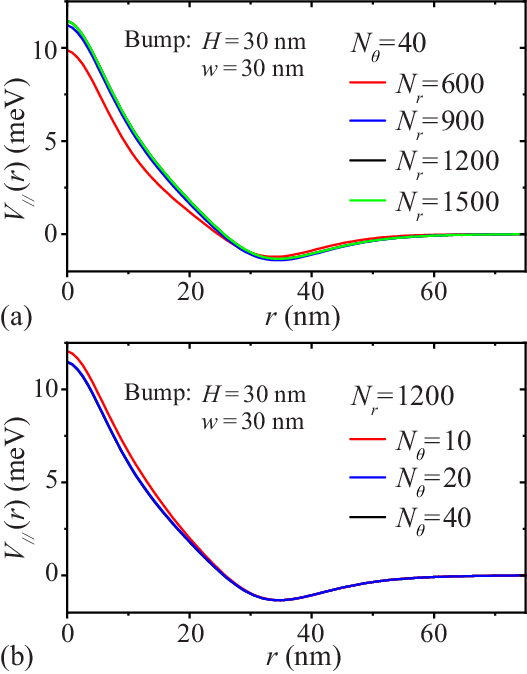}
\caption{Calculated lateral potential $V_{\parallel}(r)$ for an electron bound to a representative neon surface bump. (a) Results for different radial grid number $N_r$ with a fixed angular grid number $N_{\theta}=40$. (b) Results for different $N_{\theta}$ with a fixed $N_r=1200$.}
\label{Suppl-Fig2}
\end{figure}

To demonstrate the accuracy and convergence of our calculations, we present in Fig.~\ref{Suppl-Fig2} the computed potential energy $V_{\parallel}(r)$ associated with the lateral motion of the electron held at $h=1$~nm above a neon bump with $H=30$~nm and $w=30$~nm. We consider various values of $N{_\theta'}$ and $N_r'$ in our analysis. It is evident from the results that when $N{_\theta'}\geq20$ and $N_r'\geq900$, the computed curve of $V_{\parallel}(r)$ converges and becomes almost independent of the values of $N{_\theta'}$ and $N_r'$. After conducting an extensive practical analysis, we have selected the values $N{_\theta'}=40$ and $N_r'=1200$ as a balance between accuracy and computational efficiency.

\section{Electric force calculation}
\begin{figure}[t]
\includegraphics[width=1.0\linewidth]{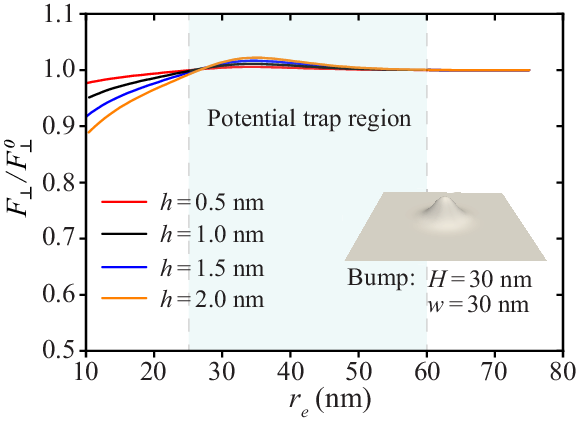}
\caption{Calculated normalized electric force $F_{\bot}/F^0_{\bot}$ acting on an electron perpendicular to the local surface of a representative neon surface bump. Curves for the electron held at different heights $h$ above the neon surface are shown.}
\label{Suppl-Fig3}
\end{figure}

Once the surface change density $\sigma(\vec{r}_s)$ is obtained, the electric force exerted on the electron can be calculated as:
\begin{equation}
\vec{F}(\vec{r}_e)=\sum_{\vec{r}_s} \frac{-e\Delta S(\vec{r}_s) \sigma(\vec{r}_s)}{4\pi\epsilon_0}\frac{\vec{r}_s-\vec{r}_e}{|\vec{r}_s-\vec{r}_e|^3}.
\label{SM-Eq2}
\end{equation}
where $\Delta S(\vec{r}_s)$ denotes the area of a surface element at $\vec{r}_s$. From this, we can determine the forces parallel $\vec{F}_{\|}(\vec{r}_e)=(\vec{F}(\vec{r}_e)\cdot \hat{n}_{\|})\hat{n}_{\|}$ and perpendicular $\vec{F}_{\bot}(\vec{r}_e)=(\vec{F}(\vec{r}_e)\cdot \hat{n}_{\bot})\hat{n}_{\bot}$ to the surface. Here, $\hat{n}_{\|}$ and $\hat{n}_{\bot}$ are unit vectors that are parallel and perpendicular to the local neon surface, respectively. The magnitudes of these forces depend on the electron's position relative to the bump and its distance $h$ from the bump surface. In Fig.~\ref{Suppl-Fig3}, we show the calculated normalized perpendicular force magnitude $F_{\bot}/F^0_{\bot}$ as a function of $r_e$ for an electron held at different heights $h$ above the neon surface, where $F^0_{\bot}$ denotes the force acting on the electron at $h$ above a flat neon surface.

It is clear that $F_{\bot}/F^0_{\bot}$ is close to one except when the electron is near the center of the surface bump at small $r_e$ values. This behavior is due to the bump surface’s curvature radius being substantially larger than $h$ away from the bump's center, effectively rendering the surface nearly flat to the electron. In particular, within the range of 25~nm~$<r_e<$~60~nm where we observe the toroidal trapping potential confining the electron's lateral motion, $F_{\bot}/F^0_{\bot}$ remains close to one regardless of $r_e$ and $h$. This observation strongly suggests that the electron's motion perpendicular to the surface closely resembles that on a flat surface. In other words, the electron is bound at about 1~nm above the neon surface, exhibiting a distribution profile $|\psi_\bot(h)|^2$ perpendicular to the surface similar to that of a flat surface.

\begin{figure}[t]
\includegraphics[width=1.0\linewidth]{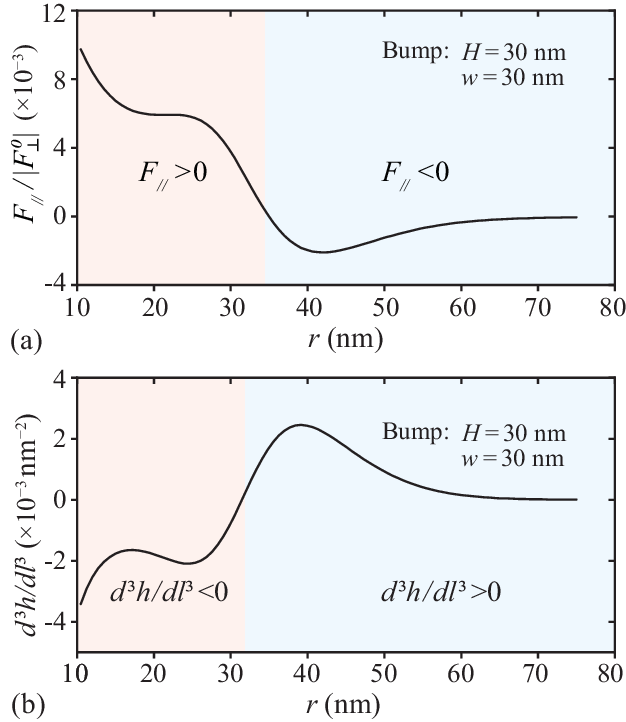}
\caption{(a) Calculated electric force $F_{\|}/F^0_{\bot}$ exerted on the electron parallel to the local surface for the same bump depicted in the preceding figure. (b) Third-order derivative $d^3h/dl^3$ depicting the variation of the electron's distance $h$ from the neon surface with respect to the coordinate $l$. For these calculations, the electron is maintained at a perpendicular distance of 1~nm above the neon surface.}
\label{Suppl-Fig4}
\end{figure}

\begin{figure}[t]
\includegraphics[width=1.0\linewidth]{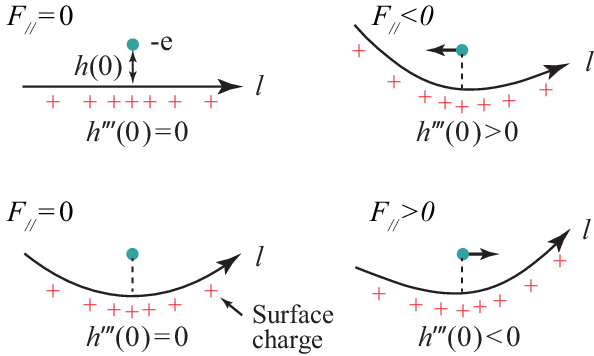}
\caption{Schematics showing how the third-order derivative $d^3h/dl^3$ correlates with the local neon surface geometry and affects the resulting parallel electric force $F_{\|}$.}
\label{Suppl-Fig5}
\end{figure}

In Fig.~\ref{Suppl-Fig4}(a), we show the calculated parallel force $F_{\|}$ exerted on an electron that is held at $h=1$~nm above the neon surface bump. Clearly, $F_{\|}$ is much smaller in magnitude as compared to $F_{\bot}$, and it changes sign from negative values at large $r_e$ to positive values at about $r_e=34$~nm. This phenomenon can be explained by considering how the distance $h(l)$ between the electron and the surface varies with the coordinate $l$, which is mapped onto the curved surface, as shown schematically in Fig.~\ref{Suppl-Fig5}. For convenience, we may set $l=0$ as the position of the electron's perpendicular projection point on the curved surface. Close to this projection point, $h(l)$ can be approximated using a Taylor expansion as $h(l)\simeq h(0)+\frac{1}{2}h''(0)l^2+\frac{1}{6}h'''(0)l^3$, where the prime denote the derivative with respect to $l$. For a flat surface or a parabolic surface with $h'''(0)=0$, the surface elements located at $-l$ and $l$ are equidistant from the electron, resulting in their induced electric forces on the electron canceling each other in the $\hat{n}_{\|}$ direction. Consequently, $F_{\|}\simeq0$ in such cases. However, when $h'''(0)>0$, the surface element at $l$ is farther from the electron compared to the surface element at $-l$, and the induced charge density at $l$ is smaller. This imbalance leads to a net electric force with a negative parallel component, i.e., $F_{\|}<0$. Conversely, when $h'''(0)<0$, a similar analysis results in $F{\|}>0$. Thus, the sign of $F_{\|}$ strongly correlates with the value of $d^3h/dl^3$ along the curved surface. In Fig.~\ref{Suppl-Fig4}(b), we show the calculated values of $d^3h/dl^3$ at various locations along the considered bump. Clearly, the curve of $F_{\|}$ is closely linked to that of $d^3h/dl^3$, and both exhibit a sign change roughly around the same location.

\section{Potential energy calculation}
\begin{figure}[t]
\includegraphics[width=1.0\linewidth]{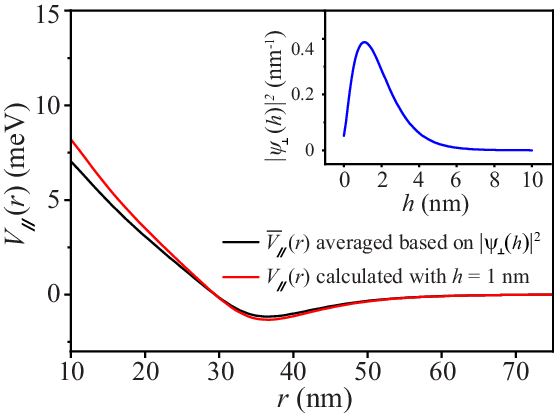}
\caption{Comparison of the lateral potential $V_{\|}(r)$ for an electron held at $h=1$~nm above a representative neon surface bump, versus the weighted potential $\overline{V_{\|}}(r)$, derived from the electron's probability $|\psi_\bot(h)|^2$ at $h$. The inset shows the profile of $|\psi_\bot(h)|^2$.}
\label{Suppl-Fig6}
\end{figure}

The potential energy associated with the electron's lateral motion on the curved neon surface is given by $V_{\|}(l)=\int_l^\infty\vec{F}_{\|}(l')\cdot d\vec{l}'$, with $l$ being the coordinate mapped along the surface, as shown in Fig.~\ref{Suppl-Fig1}. Considering the relationship between $l$ and $r$ for a Gaussian surface, we can express $V_{\|}$ in terms of $r$. Note that the probability of the electron appearing at a perpendicular distance $h$ from the neon surface is $|\psi_\bot(h)|^2$. Since the magnitude of the parallel force $F_{\|}$ depends on $h$, we need to average $V_{\|}$ weighted by $|\psi_\bot(h)|^2$ as $\overline{V_{\|}}(r)=\int V_{\|}(r; h)|\psi_\bot(h)|^2dh$. This weighted potential $\overline{V_{\|}}(r)$ is depicted in Fig.~\ref{Suppl-Fig6}. For reference, the profile of $|\psi_\bot(h)|^2$ is also provided in the figure's inset. Since $F_{\|}$ changes sign from negative at large $r$ to positive at about $r=34$~nm, the potential $\overline{V_{\|}}(r)$ exhibits a minimum at this location. A toroidal trapping potential region then emerges at 25~nm~$<r<$~60~nm. Additionally, we have also calculated $V_{\|}(r)$ while fixing the electron consistently at $h=1$~nm above the neon surface. The corresponding curve is also presented in Fig.~\ref{Suppl-Fig6}. Clearly, the $V_{\|}(r)$ curve with a fixed $h=1$~nm closely aligns with that of $\overline{V_{\|}}(r)$, especially in the toroidal trapping potential region where the electron is laterally confined. This observation implies that it is feasible to simplify our analysis by fixing the electron at $h=1$~nm, allowing for a more efficient computational examination of extensive bump geometries.

\section{Schr\"{o}dinger equation on curved surface}
The Schr\"{o}dinger equation that governs the electron's lateral motion on the neon surface is given by:
\begin{equation}
E\psi_\|(r,\theta)=-\frac{\hbar^2}{2m_e}\nabla^2\psi_\|(r,\theta)+V_{\|}(r)\psi_\|(r,\theta),
\label{SM-Eq3}
\end{equation}
where $\psi_\|(r,\theta)$ represents the electron's lateral wavefunction. On the curved two-dimensional neon surface, the term involving the Laplacian operator on the right-hand side of Eq.~(\ref{SM-Eq3}) can be evaluated as:
\begin{equation}
\nabla^2\psi_{\|}=\frac{1}{\prod_jh_j}\frac{\partial}{\partial q_i}\left(\frac{\prod_j h_j}{h^2_i} \frac{\partial \psi_{\|}}{\partial q_i}\right).
\label{SM-Eq4}
\end{equation}
In this expression, the Einstein summation convention is invoked so that repeated dummy variables indicate summation. We define the orthogonal curvilinear coordinates $q_i$ as $q_r=r$ and $q_{\theta}=\theta$. The Lam\'e coefficient, $h_i$, is given by $h_i=\lVert \frac{\partial\vec{r}_s}{\partial q_i} \rVert$. For a specific point on the 2D Gaussian bump surface, its position vector is described by: $\vec{r}=r\cos(\theta)\hat{e}_x+r\sin(\theta)\hat{e}_y+H\exp(-2r_s^2/w^2)\hat{e}_z$. Therefore, we have:
\begin{equation}
\begin{split}
&\frac{\partial\vec{r}}{\partial q_r}=\cos(\theta)\hat{e}_x+\sin(\theta)\hat{e}_y+\frac{-4r_s}{w^2}He^{-2r_s^2/w^2}\hat{e}_z\\
&\frac{\partial\vec{r}_s}{\partial q_{\theta}}=-r\sin(\theta)\hat{e}_x+r\cos(\theta)\hat{e}_y.
\end{split}
\label{SM-Eq5}
\end{equation}
As a result, $h_r=\lVert \frac{\partial\vec{r}_s}{\partial q_r} \rVert=\sqrt{1+\frac{16Hr^2}{w^2}\exp(-\frac{4r^2}{w^2})}$ and $h_{\theta}=\lVert \frac{\partial\vec{r}_s}{\partial q_{\theta}} \rVert=r$. The Schr\"{o}dinger equation can finally be expressed as:
\begin{equation}
\begin{split}
E\psi_\|(r,\theta)=&-\frac{\hbar^2}{2m_er^2}\bigg[\frac{r}{h_r}\partial_r\left(\frac{r}{h_r}\partial_r\psi_\|\right)+\partial_\theta^2\psi_\|\bigg]\\
                   &+V_{\|}(r)\psi_\|.
\end{split}
\label{SM-Eq6}
\end{equation}
Considering the axial symmetry of the potential $V_{\|}(r)$, the eigenfunction should take the form $\psi_{\|}(r,\theta)=R(r)e^{im_z\theta}$. This leads to the radial function $R(r)$ to obey:
\begin{equation}
\begin{split}
ER(r)=&-\frac{\hbar^2}{2m_er^2}\bigg[\frac{r}{h_r}\partial_r\left(\frac{r}{h_r}\partial_rR(r)\right)\\
                   &-m_z^2R(r)\bigg]+V_{\|}(r)R(r).
\end{split}
\label{SM-Eq7}
\end{equation}
The above equation is then solved using the finite difference method with a spatial resolution $\Delta r=10^{-3}w$ and with a maximum $r$ range that extends to six times the width $w$ of the bump. We have ascertained that this resolution guarantees convergence of the outcomes. The eigenfunctions $\psi^\|_{n_r,m_z}(r,\theta)=R_{n_r,m_z}(r)e^{im_z\theta}$ and the corresponding eigen-energies $E_{n_r,m_z}$ can thus be obtained.

As discussed in the main paper, the states $\psi^\|_{0,1}$ and $\psi^\|_{0,-1}$ are degenerate for a bump with perfect axial symmetry, which is unfavorable for qubit operation. This degeneracy can be lifted by an applied magnetic field. For instance, when a uniform magnetic field $\vec{B}=-B_z\hat{z}$ is applied, the eigenenergy of the electron is given by $E^{(s_z)}_{n_r,m_z}=E_{r}+\frac{\hbar^2}{2m_e}\left\langle \frac{1}{r^2}\right\rangle m_z^2-\mu_BB_z(m_z+2s_z)$, where the radial energy contribution $E_r$ is:
\begin{equation}
\begin{split}
E_r&=\int2\pi rdr\bigg[-\frac{\hbar^2R_{n_r,m_z}(r)}{2m_erh_r}\partial_r[\frac{r}{h_r}\partial_r R_{n_r,m_z}(r)]\\
   &+V_{\|}(r)R^2_{n_r,m_z}(r)+\frac{eB_z^2}{8m_e}r^2R^2_{n_r,m_z}(r)\bigg],
\end{split}
\label{Eq4}
\end{equation}
which strongly depends on $n_r$ and is only mildly affected by $m_z$. In the above expression, $\left\langle \frac{1}{r^2}\right\rangle=\int2\pi rdr[R^2_{n_r,m_z}(r)/r^2]$, $\mu_B=\frac{e\hbar}{2m_e}$ is the Bohr magneton, and $s_z=\pm\frac{1}{2}$ is the electron spin quantum number. Depends on the strength of the applied magnetic field $B_z$, the energies of the two excited states $\psi^\|_{0,1}$ and $\psi^\|_{0,-1}$ differ and can be tuned, as shown in the main paper.
